\documentstyle[12pt]{article} \topmargin=-0.4in
\input epsf.tex
\begin{document} 
{\rightline \today}
\begin{center} 
{\Huge \bf Lattice field theories with an energy current}\\
 \vskip0.1in
John Cardy\footnote{And All Souls College, Oxford}
and Peter Suranyi\footnote{On leave of absence from the University of
Cincinnati, Cincinnati, Ohio 45221}
\\ Oxford University, Theoretical Physics, 1 Keble Road, Oxford OX1 3NP\\
United
Kingdom \end{center} 
\begin{abstract}
We investigate a lattice scalar field theory in the presence of a bias
favouring the establishment of an energy current, as a model for
stationary nonequilibrium processes at low temperature in a
non-integrable system. There is a transition at a finite value of the bias
to a gapless modulated
phase which carries a classical current; however, unlike in similar,
integrable, models, quantum
effects also allow for a non-zero current at arbitrarily small bias.
The transition is second order in the
magnetically disordered phase, but is pre-empted by a first-order
transition in the ferromagnetic case, at least at the mean-field level.
 \end{abstract} 

\section{Introduction}
Finding a general framework for the description 
of nonequilibrium processes has
been a long standing problem in statistical physics.  In particular, there
is a
substantial body of literature dealing with the microscopic theory of
energy,
or heat,
conduction~\cite{first}\cite{second}\cite{third}\cite{fourth}.
Unfortunately, no clear
conclusions have been reached as a result of these investigations.
Results are
 dependent on boundary conditions and on basic assumptions concerning
the underlying process responsible for energy conduction.  

Recently, Antal, Racz, and Sasvari have proposed studying a related but
slightly different problem, namely
the microscopic description of energy conduction in quantum
systems at zero temperature~\cite{antal1}\cite{antal2}.  
Denoting the quantum hamiltonian by $H^{(0)}$ and the space integral of
the
energy current by $J$, they introduce a Lagrange multiplier $\lambda$ and
study
the ground state properties of $H^{(0)}-\lambda J$, for certain
exactly solvable one dimensional quantum spin systems. An
interesting feature of these models
is that the energy current turns on only at a
critical, nonvanishing, value of the Lagrange multiplier.  
At the critical point the system undergoes
a second order phase transition to a gapless
energy-conducting state, in which the spin
correlation function has a characteristic oscillatory behaviour,
whose amplitude is modulated by a power law fall-off.

It is of considerable interest to find out to what extent the
phenomena found in~\cite{antal1},\cite{antal2} apply to other systems, in
particular whether a phase transition to an
energy-conducting state persists in higher dimensions, and whether
integrability plays an essential role in the results.

Having departed from exactly solvable
theories, one needs a systematic approximate method to attack the
problem.  In quantum field theory the effective action~\cite{jona},
combined with the loop expansion, has been
extremely successfully in treating all kinds of problems related to
symmetry breaking.  
Standard applications of the method of effective action are usually
restricted
to the breaking of global or gauge symmetries~\cite{peskin}.  In
spontaneous
symmetry breaking of continuous symmetries in three or more dimensions a
scalar field acquires a constant vacuum expectation value.  As this
expectation value, or classical field, is constant the effective action
reduces
to the {\em effective potential}, simplifying calculations considerably.  
An
additional bonus of working with the effective action is that the inverse
correlation function is obtained as the second functional derivative of
the
effective action with respect to the classical field.  The 
definiteness of the spectral function of the propagator implies the
convexity
of the effective action.  Just like its
analogue in statistical mechanics, the Gibbs free energy, the absolute
minimum of the effective action selects the correct phase of the system. 
Although, as we shall show, 
the method of the effective action when applied to our problem 
leads to a {\em coordinate dependent} classical field and, thus,
spontaneous
breaking of translation invariance, we expect that the effective action is
still minimised by the correct classical field configuration. 

\begin{figure}[htbp]
\epsfxsize=9cm
\centerline{
\epsfbox{
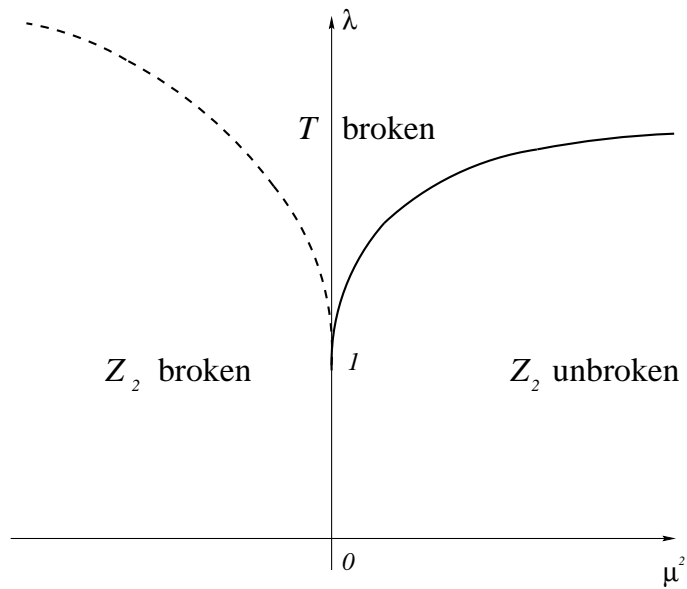}}
\caption{\em Schematic phase diagram at $D>1$. $T$ signifies translation
symmetry. 
 The dashed line represents a first order transition.}
\label{figphasediag}
\end{figure}

The model we consider in this paper is a continuous spin version, in $D$
space dimensions, of the
transverse Ising model discussed in \cite{antal1} for $D=1$.
In the absence of a current bias, the model has a phase transition from
a disordered state to one in which in which the ${\rm Z}_2$ symmetry
is spontaneously broken, in the universality class of the
$(D+1)$-dimensional classical Ising model. The phase diagram in the
presence of the current bias $\lambda$ is shown schematically in 
Fig.~\ref{figphasediag}.
Qualitatively it is quite similar to that found in
\cite{antal1}. However, there are important differences.
At a finite value of $\lambda$ there is a transition to a
situation in which, for $D>1$,
the ground state is no longer invariant under lattice translations. In
this phase there is a characteristic wave number $p_0$ such that Fourier
components of the field which are an integer multiple of this acquire a
vacuum expectation value. For $\mu^2>0$, only the fundamental
frequency is important near the transition,
 and its amplitude vanishes like
$(\lambda-\lambda_c)^{1/2}$. This classical spin density wave 
supports a non-zero energy current. There are Goldstone modes
corresponding to this breaking of translational symmetry, and for $D=1$
these actually destroy the long range order.  The phase transition is
then induced by the condensation of vortices, just like in the $D=1$
planar
rotor model.  The Goldstone modes lead to a power-law
modulation of the amplitude of the oscillatory spin correlation
function.   Even for $D=1$, this state
with only quasi-long range order is still capable of supporting a
classical current. 

However, our model differs from those considered 
in~\cite{antal1},\cite{antal2} in that, due to quantum effects, the
current
is also nonvanishing in the region $0<\lambda<\lambda_c$. This may
be simply understood as follows. In~\cite{antal1},\cite{antal2} the
space integral $J$ of the current happens to commute with the
hamiltonian $H^{(0)}$. This means, in particular, that the vacuum state
of  $H^{(0)}$ is also an eigenstate of $H=H^{(0)}-\lambda J$. In the
general case where the spectrum of  $H^{(0)}$ has a gap, we expect the
vacuum state of $H$ to depend analytically on $\lambda$, and therefore
to remain unchanged at least for some finite neighbourhood of
$\lambda=0$. In this region, the current will continue to vanish, and,
indeed, all the equal-time correlations will be identical to the case
when $\lambda=0$. The onset of the current then occurs at the first
nonanalyticity, $\lambda=\lambda_c$. In general, and in our model in
particular, there is no a priori reason for  $H^{(0)}$ and $J$ to
commute, and the structure of the vacuum may change as soon as 
$\lambda\not=0$, allowing the current to flow. Interesting enough, we
find that this happens due to quantum processes which first arise
only at $O(\hbar^3)$, so that such effects might be very small in a real
system.

Most of our analysis applies to the case when $\lambda$ is switched on
in the disordered, symmetric phase of the continuous spin Ising model.
However, a very similar picture seems to hold in the ferromagnetic
phase. There is a transition at a finite value of $\lambda$ to
a non-translational invariant state (for $D>1$) which carries a
classical energy current. However, in our particular model, in weak
coupling, it can be shown that the expected second order transition is
preempted by a first order one to a modulated
state in which the higher harmonics
are important. 

The outline of this paper is as follows.
In the next section, after setting up an appropriate hamiltonian which 
captures the physics we want to describe,
we calculate the effective
action at tree level.  
Sec.~3. discusses the theory and its quantum corrections
in the ferromagnetically disordered
phase, first at small values of the bias, then at the critical value,
and finally in the modulated phase. In all three cases we evaluate the
classical current and its lowest order quantum corrections.
In Sec.~4 we pay particular attention to the Goldstone modes which
appear in the modulated phase for $D>1$ and discuss how these
dramatically modify the physics for $D=1$.  The next section 
describes our (limited) results for the ferromagnetically ordered phase,
where there appears to be a first-order transition to the modulated
phase at mean-field level. Finally, in Sec.~6 we summarise our results,
contrast them with those found in the integrable $D=1$ example of
Ref.~\cite{antal1}, and connect them with those in other different, but
mathematically similar, models of modulated phases.

\section{A lattice model} 

Our first aim is to select a suitable field theory hamiltonian 
$H^{(0)}$ whose
long-distance behaviour is expected to be in the same universality
class as the transverse Ising model whose $D=1$ version was
discussed in \cite{antal1}. In the
absence of any Lagrange multiplier, it is well known that the continuum
relativistic $\phi^4$ field theory, with a suitable momentum cut-off,
has this property, and that it undergoes a zero-temperature phase
transition in the same universality class as the classical Ising model
in $D+1$ space dimensions.

However, it turns out that such a relativistic continuum theory cannot
exhibit the physics we are trying to study at non-zero $\lambda$. This
is because the energy current is given by the components $T_{0i}$ of
the energy-momentum tensor, which, in a relativistic theory, are equal
to those of the momentum density $T_{i0}$. The current $J$ is therefore
nothing
but the total momentum $P$, which commutes with $H^{(0)}$. Any
simultaneous eigenstate has, for large $P$, an energy $E\sim P$.
{}From the convexity of the relativistic dispersion relation it is
easy to see that the ground state of $H^{(0)}-\lambda P$ is the usual
vacuum for $\lambda<1$, while for $\lambda>1$ this operator is unbounded
from below, so that the theory does not exist. 

Clearly it is necessary to modify the dispersion relation, and the
natural way to do this is to consider the same field theory on a
lattice, since this also provides an ultraviolet regulator.
The hamiltonian of the $D$-dimensional lattice model that we
consider is thus 
\begin{equation} H^{(0)} = \frac{1}{2}\sum_{\mbox{\boldmath
$r$}}\mbox{\Huge\{}\Pi^2_{\mbox{\boldmath $r$}}
+\mu^2\phi^2_{\mbox{\boldmath
$r$}}+\sum_\alpha^D [\phi_{\mbox{\boldmath $r$}}-\phi_{\mbox{\boldmath
$r$}-\mbox{\boldmath $e$}_\alpha}]^2 +\frac{g}{12}\phi^4_{\mbox{\boldmath
$r$}}\mbox{\Huge\} }, 
 \label{lattice} 
\end{equation}  
where summation is taken over
the vertices of a $D$-dimensional cubic lattice and
$\mbox{\boldmath$e$}_\alpha$
are unit (lattice) vectors. The lattice constant is taken to be unity. The
canonical momentum, $\Pi$, is introduced and the time coordinate is kept
continuous to facilitate transformation to the lagrangian formalism later. 

Let us calculate the energy current $j_{\mbox{\boldmath$r$},\alpha}^E$
of this hamiltonian. This satisfies
 \begin{equation}
i[H^{(0)},H^{(0)}_{\mbox{\boldmath
$r$}}]=\nabla_\alpha j_{\mbox{\boldmath$r$},\alpha}^E\, ,
\label{commutator} 
\end{equation}
where $\nabla_\alpha$ denotes the lattice divergence. The solution of
(\ref{commutator}) for the space integral of the energy current, the only
quantity of interest for us, is unique.
 
After a simple calculation we find
from (\ref{lattice},\ref{commutator})
 \begin{equation}
 j^E_{\mbox{\boldmath $r$},\alpha}=\frac{1}{2}\left[(\Pi_{\mbox{\boldmath
$r$}}+\Pi_{\mbox{\boldmath $r$}-\mbox{\boldmath
$e$}_\alpha})(\phi_{\mbox{\boldmath $r$}}-\phi_{\mbox{\boldmath
$r$}-\mbox{\boldmath $e$}_\alpha})\right] .
\label{energy_current} 
\end{equation}
Note that this is hermitian but is odd under time-reversal and parity.

Now we modify our lattice hamiltonian by the addition of  the space
integral of a component of the energy current, $\sum_{\mbox{\boldmath
$r$}}j^E_{\mbox{\boldmath $r$},1}$, multiplied by a Lagrange multiplier,
to get  
\begin{equation} H =
\frac{1}{2}\sum_{\mbox{\boldmath $r$}}\mbox{\Huge\{}\Pi^2_{\mbox{\boldmath
$r$}}
+\mu^2\phi^2_{\mbox{\boldmath $r$}}+\sum_\alpha^D \,[\phi_{\mbox{\boldmath
$r$}}-\phi_{\mbox{\boldmath $r$}-\mbox{\boldmath $e$}_\alpha}]^2 +
\frac{g}{12}\phi^4_{\mbox{\boldmath $r$}}+\frac{\lambda}{2}\,
j^E_{\mbox{\boldmath $r$},1}\mbox{\Huge\} },  
\label{latticeh2}  
\end{equation}
Note that we have selected the first coordinate as the direction of the
energy
current.

 The  term $\sum_{\mbox{\boldmath $r$}}j^E_{\mbox{\boldmath $r$},1}$ does
not commute with $H^{(0)}$, due to the presence of the $\phi^4$
interaction term.  In fact the commutator is proportional to
\begin{equation}
\sum_{\mbox{\boldmath $r$}}\phi^3_{\mbox{\boldmath $r$}}
[\phi_{\mbox{\boldmath $r$}+\mbox{\boldmath $e$}_1}-
\phi_{\mbox{\boldmath $r$}-\mbox{\boldmath $e$}_1}]
\end{equation}
Note that in the naive continuum limit this is proportional to
$\phi^3\partial_1\phi$, which is a total derivative, but when higher
order derivatives are taken into account it has a non-vanishing effect.

\begin{figure}[htbp]
\epsfxsize=9cm
\centerline{\epsfbox{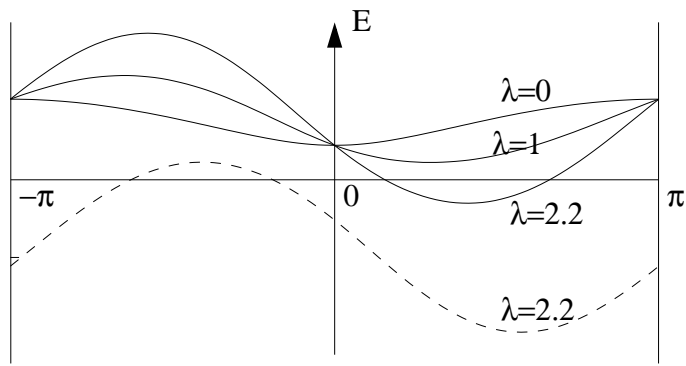}}
\caption{Dispersion relations at $\lambda=0$, $\lambda=1<\lambda_c$,
$\lambda=2.2>\lambda_c$ (solid lines), and for the `negative energy
branch' at $\lambda=2.2$ (dashed line).}
\label{fig1pd}
\end{figure}

As we emphasised in the Introduction, the main advantage of our field
theoretic
model is that we are able to utilise the lagrangian formalism.  
However, it is instructive first to examine the hamiltonian
(\ref{latticeh2}) in the
limit $g=0$ when the self-interaction is neglected. In that case $H$
may be diagonalised in terms of momentum-space annihilation and creation
operators, with the result
\begin{equation}
\label{hfree}
H=\int{d^D\!p\over(2\pi)^D}\left(2\sum_{\alpha=1}^D(1-\cos p_\alpha)
+\mu^2)^{1/2}-\lambda\sin p_1\right)a^{\dag}(p)a(p),
\end{equation}
where the integral is over the first Brillouin zone. For $\mu^2>0$
and small $\lambda$, the effect is only to tilt slightly the usual
one-particle dispersion relation (solid line), as illustrated by  the
dotted line of Fig.~\ref{fig1pd}.
The ground state is however still the original vacuum. However, there
is a critical value of $\lambda>0$ for which the dispersion curve touches
the zero-energy axis at some positive $p_1=p_1^{(0)}$, and beyond this
value there are negative energy single particle states
(dashed line of Fig.~\ref{fig1pd}), which will carry
a nonzero energy current. Unlike the case considered in \cite{antal1},
however, where the elementary excitations were fermions, the particles
will now try to Bose condense into the lowest energy single particle
state.
Since the number of such quanta
is unbounded, the energy will diverge
to minus infinity, an unphysical result. Of course, this picture will
change in the presence of the repulsive $\phi^4$ interaction, since
this will limit the total number of quanta. However, the interaction
will also introduce scattering between quasiparticles of different 
momenta, and in particular will introduce virtual processes involving
the branch of the dispersion relation found by taking the other
sign of the square root, indicated by the broken line in
Fig.~\ref{fig1pd}. For $\lambda$ greater than its critical value the
energies of some of these states begin to overlap with those of the
original single-particle states, indicating a complete
breakdown of the quasi-particle picture. 

For this reason, among others, we have found it more useful to analyse
the interacting theory through the path integral and the effective
action formalism.
First we need to
calculate the lagrangian. We use the canonical equation to find
\begin{equation}
\label{phidot}
\dot\phi_{\mbox{\boldmath $r$}}=\Pi_{\mbox{\boldmath
$r$}}+\frac{\lambda}{2} [\phi_{\mbox{\boldmath $r$}+\mbox{\boldmath
$e$}_1}-\phi_{\mbox{\boldmath $r$}-\mbox{\boldmath $e$}_1} ].
\end{equation}
Then the lagrangian,
 \begin{equation}
L=\sum_{\mbox{\boldmath $r$}}\Pi_{\mbox{\boldmath $r$}}\dot
\phi_{\mbox{\boldmath $r$}} - H,
\end{equation} 
takes the form 
\begin{eqnarray} L &=&
\frac{1}{2}\mbox{\Huge\{}\sum_{\mbox{\boldmath
$r$}}[\dot\phi_{\mbox{\boldmath
$r$}}-\frac{\lambda}{2} {\Large(}\phi_{\mbox{\boldmath
$r$}+\mbox{\boldmath
$e$}_1}-\phi_{\mbox{\boldmath $r$}-\mbox{\boldmath $e$}_1}{\Large)}]^2
-\mu^2\phi^2_{\mbox{\boldmath $r$}}\nonumber \\ &- &\sum_\alpha^D
[\phi_{\mbox{\boldmath $r$}}-\phi_{\mbox{\boldmath $r$}-\mbox{\boldmath
$e$}_\alpha}]^2- \frac{g}{12}\phi^4_{\mbox{\boldmath $r$}}\mbox{\Huge\} },
\label{lagrange} 
\end{eqnarray}

In the lowest order of the
loop expansion the effective action is identical to the classical
action.
Then, at tree level, the classical field minimises the classical action. 
As we have argued above, for $\lambda$ larger than a critical value we
expect to find a macroscopic number of particles with momentum
$\approx p_1^{(0)}$ in the ground state. This means that the classical
field which minimises the action will be space-dependent. {}From the point
of view of the path integral, there is also no reason to exclude a
time-dependence as well. We shall assume that, since we are considering
a stationary process, that this is not the case. Further justification
for this may be found by considering the euclidean action, whose
minimum is guaranteed to give the lowest energy state. This is found
from (\ref{lagrange}) by changing the sign of all the terms and also
letting $\dot\phi\to i\dot\phi$. In general, then, euclidean saddle
point solutions will be complex, unless they are time-independent.
Since we are seeking a ground state with a real field, we should
exclude complex solutions.

If we allowed the classical field to depend on time the solution of the
extremisation problem would become far from unique. In fact, one could
find a
solution at all values of the parameters, $\lambda$ and $\mu^2$. 
Furthermore, on a more practical level, we would completely lose the
guiding principle  of finding the classical
field by minimising the effective action.   One can see this problem
arising
already at the classical level. If the classical field would depend on
time,
then its Fourier transform would depend on `energy' and the kinetic part
of the
lagrangian would become indefinite.  Looking at this question from a
slightly
different angle, remember  that the second functional derivative of the
effective action, or at tree level, classical action, is the inverse
propagator.  The inverse of this, the propagator has a spectral
representation
with a positive definite weight function and non-negative threshold.
Thus, at
any space or timelike four-momentum (or generalised four momentum for
lattices)
the propagator is positive.   In momentum representation this insures the
definiteness of the inverse propagator, as well.  However, if the energy
component is nonvanishing then the spectral representation does not
insure the definiteness of the propagator and there is no minimum
principle for
the effective action.

Note that, as we shall discuss in detail later, the fact that the
classical field is time-independent does not preclude the ground state
from carrying a current. This is because, as may be seen from
(\ref{energy_current},\ref{phidot}),
if $\phi$ is space-dependent then $\dot\phi$ may vanish without $\Pi$,
and therefore the classical current, vanishing. {}From the hamiltonian
perspective, this is related to the fact that the `time' appearing in
the lagrangian (\ref{lagrange}) is not in fact the true time, but rather
the conjugate quantity to the effective hamiltonian $H=H^{(0)}-\lambda
J$, whose ground state properties we are trying to find. The real
time-dependence of the system is still generated by the original
hamiltonian $H^{(0)}$. Thus, in the Heisenberg picture, having a
`time'-independent classical field which commutes with $H$
in fact corresponds to a true time dependence
$\dot\phi=-i\lambda[J,\phi]$ - thus, as expected, local inhomogeneities
in $\phi$ are transported along by the current in this state.

\section{Energy current in the disordered phase $\mu^2>0$.}

The minimisation of the classical action\footnote{For convenience, we
will use the
negative of the effective action that will have a maximum, rather that a
minimum at the correct classical field.} on
functions independent of time leads to the following equation for
classical
field $\phi^{\rm cl}_{\mbox{\boldmath $r$}}$ that will be henceforth
denoted by
$\phi_{\mbox{\boldmath $r$}}$:
 \begin{eqnarray} i\frac{\delta
\Gamma_{\rm eff}}{ \delta \phi_{\mbox{\boldmath
$r$}}}&=&\frac{\lambda^2}{4}\left[2\phi_{\mbox{\boldmath
$r$}}-\phi_{\mbox{\boldmath $r$}-2\mbox{\boldmath
$e$}_1}-\phi_{\mbox{\boldmath
$r$}+2\mbox{\boldmath $e$}_1}\right]\nonumber \\
&-&\sum_\alpha^D\left[2\phi_{\mbox{\boldmath $r$}}-\phi_{\mbox{\boldmath
$r$}-\mbox{\boldmath $e$}_\alpha}-\phi_{\mbox{\boldmath
$r$}+\mbox{\boldmath
$e$}_\alpha}\right] -\mu^2\phi_{\mbox{\boldmath $r$}}
-\frac{g}{6}[\phi_{\mbox{\boldmath $r$}} ]^3=0.  \label{equ} 
\end{eqnarray}

Introduce the Fourier decomposition 
of the
time dependent classical field $\phi_{\mbox{\boldmath $r$}}$
\begin{equation} \phi_{\mbox{\boldmath
$r$}}=\frac{1}{(2\pi)^{(D+1)/2}}\int_{-\pi}^\pi d^Dp\,\int_0^\infty
dE\left[
\psi(\mbox{\boldmath$p$},E)e^{i\mbox{\boldmath$p$}\cdot
\mbox{\boldmath$r$}-iEt}
+ {\rm h.\, c.}\right]. \label{fourier} \end{equation}
and let $\phi(\mbox{\boldmath$p$},E) =
\psi(\mbox{\boldmath$p$},E) +
\psi^\dagger(-\mbox{\boldmath$p$},-E).$\footnote{This implies the
relations
$\phi(\mbox{\boldmath$p$},E) = \psi(\mbox{\boldmath$p$},E)$ for $E>0$
and 
$\phi(\mbox{\boldmath$p$},E) = \psi^\dagger(-\mbox{\boldmath$p$},-E)$ if
$E<0$.}

A time-dependent solution for the classical field then has the form
setting $\phi(\mbox{\boldmath$p$},E)
={2\pi}\phi(\mbox{\boldmath$p$})\delta(E)$ (\ref{equ})  
where $\phi(\mbox{\boldmath$p$})$ satisfies

\begin{equation} i\frac{\delta \Gamma_{\rm
eff}}{\delta\phi(\mbox{\boldmath$p$})
}=0=D^{-1}_0(\mbox{\boldmath$p$},0)\phi(
\mbox{\boldmath$p$})-\frac{g}{6}[\phi^3](\mbox{\boldmath$p$}), 
\label{mom_space} 
\end{equation}
and we have introduced the notation
\begin{equation}
D^{-1}_0(\mbox{\boldmath$p$},E)=(E+\lambda \sin p_1)^2-
2\sum_\alpha^D(1-\cos p_\alpha) -\mu^2
\label{unperturbed}
\end{equation}
and
\begin{equation}
[\phi^k](\mbox{\boldmath$p$})=\frac{1}{(2\pi)^{D}}\int d^Dp^1...d^Dp^k
\delta(\mbox{\boldmath$p$}-\mbox{\boldmath$p$}^1-...
-\mbox{\boldmath$p$}^k)
\phi( \mbox{\boldmath$p$}^1)...\phi( \mbox{\boldmath$p$}^k).
\label{folding}
\end{equation}

The tree level expression for the inverse propagator is 
\begin{eqnarray}
i\frac{\delta^2 \Gamma_{\rm eff}}{\delta
\phi(\mbox{\boldmath$p$},E)\delta\phi(-\mbox{\boldmath$p$}',-E')}&\sim
&\delta(E-E')\left\{\delta( \mbox{\boldmath$p$}-\mbox{\boldmath$p$}')
D^{-1}_0(\mbox{\boldmath$p$},E)
-\frac{g}{2} [\phi^2](\mbox{\boldmath$p$}-\mbox{\boldmath$p$}')\right\}.
\label{prop-old} 
\end{eqnarray}

Note that $D^{-1}_0(\mbox{\boldmath$p$},E)$ vanishes on the single
particle dispersion curve discussed earlier, as well as on the
`negative-energy' branch. 
The coefficient  
$
D^{-1}_0(\mbox{\boldmath$p$},0)
$
 in (\ref{mom_space}) can be easily analysed.  It may have several
extrema.
Besides the one at $\mbox{\boldmath$p$}=0$, provided that
$\lambda>1$, there are two at $p_1=\pm p_1^{(0)},$ $p_\alpha=0$, if
$\alpha\not=1$, where $\cos p_1^{(0)}=1/\lambda^2$. At $\lambda>1$  the
latter ones are the only maxima. The value of $D^{-1}_0$ at these maxima
is
\begin{equation} D^{-1}_0(\mbox{\boldmath$p$}^{(0)}) =
(\lambda-1/\lambda)^2-\mu^2.  \label{max}  \end{equation}
 Then for every real value of the mass gap, $\mu,$ there is a
$\lambda_c>1$, such
that the maximum of this coefficient is exactly zero, 
\begin{equation}
\lambda_c=\frac{1}{2} [\mu+\sqrt{\mu^2+4}] 
\label{critical} 
\end{equation}

Note that the convexity of the effective action requires that after
setting
$E=0$ (\ref{prop-old}) is negative at the appropriate classical field. 
Indeed, when $ D^{-1}_0(\mbox{\boldmath$p$})>0$  one can find a
nonvanishing
solution of (\ref{mom_space}). In fact, this nonvanishing solution will
provide
the minimum of the action in contrast to the symmetric solution, $\phi=0$.
We
expect that if the vacuum expectation value is independent of time then
the
propagator (the second derivative of the action) is negative definite.
Then the
effective action is a convex function of the classical field and the
classical field should be chosen to minimise it. 

\subsection{The symmetric phase}
Suppose first that  $\lambda<\lambda_c$.  Then
$D^{-1}_0(\mbox{\boldmath$p$}^{(0)})<0$ and  the only real solution of
(\ref{mom_space}) is $\phi_{\mbox{\boldmath$p$}}=0$.  Consequently,  the
symmetry is unbroken.  The tree level inverse
propagator becomes   \begin{equation}
D_0^{-1}(\mbox{\boldmath$p$},E)=\Delta_0^{-1}(\mbox{\boldmath$p$},E+\lambda
\sin
p_1), \label{unbroken_prop}  \end{equation} 
where $\Delta_0$ is the propagator in the
absence of the perturbation. Then the only change in
the perturbed theory is the change of the energy-momentum dispersion
relation to
\[
E=-\lambda\sin p_1 + \sqrt{\mu^2+2\sum_\alpha (1-\cos p_\alpha)}.
\]
Though this relation dips near
$\mbox{\boldmath$p$}=\mbox{\boldmath$p$}^{(0)}$, the energy is always
positive.
In other words, the energy gap is nonvanishing.  Perturbative corrections
are
not expected to change this result, but they may change the location of 
the critical point.

Although the propagator is modified when $\lambda\not=0$, in fact the
effect on the ground state properties is rather subtle. 
For example, the equal-time spin correlation function in the absence of
loop corrections entails integrating $D_0(\mbox{\boldmath$p$},E)$
over all $E$, and therefore the correction term $\lambda\sin p_1$
is simply shifted away. Does this result survive the addition of loop
corrections? This may be examined by making a time-dependent
transformation
on the Fourier components of the field 
\begin{equation}
\phi(\mbox{\boldmath$p$},t)\to \phi(\mbox{\boldmath$p$},t)\,
e^{it\lambda\sin p_1},
\end{equation}
which shifts away the $\lambda\sin p_1$ term in the bare propagator.
The interaction now has the form
\begin{equation}
\label{shiftedint}
\int dt \prod_{n=1}^4[d^D\!\mbox{\boldmath$p$}^{(n)}
\phi(\mbox{\boldmath$p$}^{(n)},t)]\,e^{it\lambda\sum_{n=1}^4\sin
p^{(n)}_1}
\delta(\sum_{n=1}^4\mbox{\boldmath$p$}^{(n)}).
\end{equation}
{}From this it can be seen that if we approximate $\sin p_1$ by $p_1$,
the $t$-dependence disappears and the result is the same as in the
theory with $\lambda=0$. This is clearly connected with the fact that,
to lowest order in the derivative expansion, the current does commute
with the interaction. 

\begin{figure}[htbp]
\epsfxsize=9cm
\centerline{
\epsfbox{
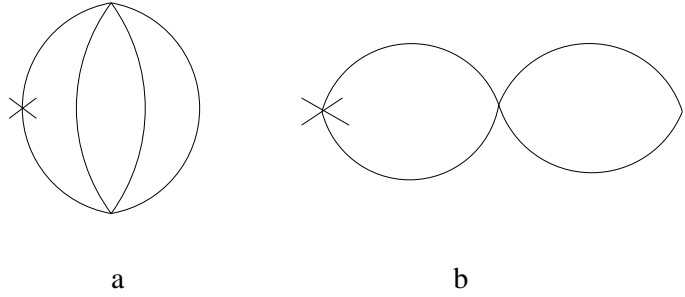}}
\caption{\em Diagrams contributing to the expectation value of the energy
current:
a) Lowest order nonvanishing diagram, 
b) Vanishing two loop diagram. The cross corresponds to a bare energy
current insertion.}
\label{figcurrent}
\end{figure}

The subtlety of this modification may be seen, for example, in the
computation of the expectation value of the
current in this phase. {}From (\ref{energy_current},
\ref{phidot}) this is
\begin{equation}
\langle j^E_1\rangle=
\int dEd^D\!\mbox{\boldmath$p$}(E+\lambda\sin p_1)\, \lambda\,\sin p_1
\langle\phi(\mbox{\boldmath$p$},E)\phi(-\mbox{\boldmath$p$},-E)\rangle.
\end{equation}
If we make the above shift $E\to E+\lambda\sin p_1$, and we compute
the correlator in the absence of any interaction, it is easy to see that
this vanishes through the symmetry $E\to-E$ (or $p_1\to-p_1$).
Indeed, this will remain true even when the interaction is included, if
we can ignore the momentum-dependent phase factors in (\ref{shiftedint}),
since these symmetries then continue to hold. But the phase factors
couple these two symmetries together, and, since the current is even
under the joint reflection of $E$ and $p_1$ it does not necessarily
vanish when loop corrections are included. However, it turns out that
the first non-zero contribution first occurs at three loops, due to the
diagram shown in Fig.~\ref{figcurrent}a. The one- and two-loop diagrams
still vanish. For example, at the internal vertex in
the diagram shown in Fig.~\ref{figcurrent}b
the incoming momenta are $(\mbox{\boldmath$p$},-\mbox{\boldmath$p$},
\mbox{\boldmath$p$}',-\mbox{\boldmath$p$}')$ so the phase 
$\sum_{n=1}^4\sin \mbox{\boldmath$p$}_n$ still vanishes.
We conclude that the first non-vanishing contribution to the current
in this phase occurs at three loops, and so is $O(g^2\hbar^3)$.

\subsection{ The critical theory} 
At the critical point $\lambda=\lambda_c$, the classical field  still
vanishes at the minimum of the action. Therefore the tree level propagator
has
the form 
\begin{equation} D(\mbox{\boldmath$p$},E)=\frac{1}{(E+\lambda_c \sin
p_1)^2-2\sum_\alpha (1-\cos p_\alpha) - \mu^2} 
\label{prop0} 
\end{equation} 
It is possible to make the same transformation as above and shift $E$
in the bare propagator. However, this obscures the fact that there
is now no energy gap, and the theory may develop
infrared singularities close to $E=0$
which modify the large distance behaviour of the
correlation functions. 
This can be seen if we expand the denominator of the propagator around
$\mbox{\boldmath$p$}=\pm\mbox{\boldmath$p$}^{(0)}$ 
and 
$E=0$.  
The constant term
vanishes at the critical point.  The term linear in 
$\mbox{\boldmath$p$}$ 
also vanishes, as 
$\pm\mbox{\boldmath$p$}^{(0)}$
 were defined as  locations of 
maxima of the inverse propagator.  Then, keeping leading order terms in
$E$
and $\mbox{\boldmath$p$}\mp \mbox{\boldmath$p$}^{(0)}$ only, we obtain 
\begin{equation}
D(\mbox{\boldmath$p$},E)\simeq\frac{1}{\pm 2 E\lambda_c \sin
p_1^{(0)}-\lambda_c^2\sin^2 p_1^{(0)}(p_1\mp
p_1^{(0)})^2-\sum_{\alpha\not=1}p^2_\alpha},  \label{prop1}
\end{equation}
 where  
\begin{equation} 
p_1^{(0)}=\cos^{-1}\frac{1}{\lambda_c^2}=\cos^{-1}\left(1-\mu\frac{\sqrt{\mu^2+4}-\mu}{2}\right)>0. 
\label{cosine} 
\end{equation} 
 The two signs in (\ref{prop1}) correspond to  poles that are separated
from in
other by a finite distance.
These two poles lead to two propagators of
states which, at least in the absence of interactions, do not mix.  
After rescaling 
$E$, 
as 
$2E\lambda_c|\sin p_1^{(0)}|\rightarrow E$, 
and  shifting and rescaling
$p_1$ 
as 
$\lambda_c \sin p_1^{(0)} ( p_1-p_1^{(0)})\rightarrow p_1$ 
we obtain the
following expression, which can be regarded as
\def\Abar{{\bar A}}
propagators for a `particle' $A$ and for an `antiparticle' $\Abar$,
both, however,
non-relativistic:  
\begin{equation}
D^\pm_F(\mbox{\boldmath$p$},E)\simeq\frac{1}{\pm E-
\mbox{\boldmath$p$}^2+i\epsilon}. 
\label{prop3} 
\end{equation}

When the interactions are included, 
in terms of the non-relativistic excitations
above, the terms allowed by overall momentum conservation are
the scattering processes
$AA\to AA$, $\Abar\Abar\to\Abar\Abar$, and also $AA\Abar\Abar\to0$
and its inverse. Without these last two processes, the theory is
identical to that of non-relativistic particles with a local 2-body
repulsive interaction. Such as theory does have infrared singularities
below $D=2$, but they do not affect the propagator, since it is easy to
see that there are in fact no loop corrections. For the same reason the
quantum corrections to the current all vanish.

This is no longer true in the presence of the last two vacuum processes,
which indeed lead, at two loops, to corrections to the current
corresponding to those illustrated, for $\mu^2>0$, in 
Fig.~\ref{figcurrent}a. However, because of the way the momentum 
flows through these diagrams, it can be seen that they are not in fact
infrared singular. Thus we expect that the quantum corrections to the
current are not singular, leaving the singularity in
classical current, to be studied in the next section.

\subsection{The broken phase}

When $\lambda $ is increased beyond its critical value, $\lambda_c$, then
the maximum of the inverse propagator becomes positive,  
$D^{-1}_0(\mbox{\boldmath$p$}^{(0)})>0$. Then a
symmetry breaking solution of
(\ref{mom_space}) exists, along with the symmetric solution,
$\phi(\mbox{\boldmath$p$})=0$.  
As the nonzero solution minimises the effective
action we must choose it over the symmetric one.
 
It is not a priori clear whether the relevant solution of
(\ref{mom_space}) is a
classical field with a sharp spectrum (i.e. delta function in momentum
space) or at any given $\lambda$  the spectrum of the classical field
contains all the momenta for which  
$D^{-1}(\mbox{\boldmath$p$})>0$.  
Examining the second functional derivative resolves this
problem.  On one hand, the convexity of the action demands  that  the
right hand side of (\ref{prop-old}) is
negative even in the neighbourhood of
$\mbox{\boldmath$p$}=\mbox{\boldmath$p$}^{(0)}$.  
On the other hand, at $\lambda>\lambda_c$ the multiplier of the momentum
conservation delta function (that is  $D^{-1}(p)$) is positive in the
neighbourhood of $p^{(0)}$. The negativity of the second variation can
only
be secured if
 the last term
of (\ref{mom_space}) also contains a term proportional to
$\delta(\mbox{\boldmath$p$}-\mbox{\boldmath$p$}')$.  
In turn, that is possible
only if 
$\phi(\mbox{\boldmath$p$})$ 
itself contains a term proportional to a
delta function (sharp spectrum), as well. Suppose that the delta function
fixes the momentum at $p=k_0$, or in other words,  $\phi\sim
\delta(p-k_0)$.  Then it is easy to see that the minimum of the action, $
S_{\rm min}\sim -[D^{-1}(k_0)]^2$.  This is minimised, as a function of
$k_0$ if $k_0=\pm p^{(0)}$.  In other words,  the delta function in the
classical field must  set
the momentum equal to 
$\pm\mbox{\boldmath$p$}^{(0)}$.

Since $\phi(x,t)$ is real at
$E=0$ 
both  ``negative'' and ``positive'' frequency terms contribute.
Therefore, we use a symmetric  ansatz
\begin{equation} \phi(\mbox{\boldmath$p$})=\phi_0(
\mbox{\boldmath$p$}) +\Delta\phi(\mbox{\boldmath$p$}), 
\label{ansatz}
\end{equation}
 where 
\begin{equation} \phi_0( \mbox{\boldmath$p$}) =\rho\left[ e^{i\theta_0} 
\ \delta(\mbox{\boldmath$p$}-\mbox{\boldmath$p$}^{(0)})+  e^{-i\theta_0}
\
\delta(\mbox{\boldmath$p$}+\mbox{\boldmath$p$}^{(0)})\right].
\label{ansatz3}
\end{equation}

In (\ref{ansatz3}) $\rho$ is a real amplitude, and 
$\theta_0$ 
is an arbitrary real phase. As we shall see  in the next section the
arbitrary phase is related to the presence of
a Goldstone boson in the broken phase. 

Near the phase transition the term $\Delta\phi(\mbox{\boldmath$p$})$ is
much smaller than the leading term, $\phi_0(\mbox{\boldmath$p$})$.  The
self-consistency of this requirement will become obvious below.

Upon substituting (\ref{ansatz}) into condition (\ref{mom_space}) 
we obtain 
\begin{eqnarray} 
0&=&
\left[-\mu^2+\left(\lambda-\frac{1}{\lambda}\right)^2-\frac{g}{2(2\pi)^D}\rho^2\right]\phi_0(
\mbox{\boldmath$p$})\nonumber\\ &+& \left[\lambda^2 \sin^2p_1
-2\sum_\alpha\left(1-\cos p_\alpha \right) - \mu^2 
\right]\Delta\phi(\mbox{\boldmath$p$}) \nonumber \\
&-&\frac{g}{6(2\pi)^D}\rho^3
\left[ e^{3i\theta_0}
\delta(\mbox{\boldmath$p$}-3\mbox{\boldmath$p$}^{(0)})+ 
e^{-3i\theta_0}
\delta(\mbox{\boldmath$p$}+3\mbox{\boldmath$p$}^{(0)})\right]
+O(\rho^2\Delta\phi(\mbox{\boldmath$p$})). 
\label{subst2} 
\end{eqnarray}

Consequently, the leading order term ($O(\lambda-\lambda_c)$),
proportional to $ \delta(\mbox{\boldmath$p$}\pm\mbox{\boldmath$p$}^{(0)})$
of (\ref{subst2}) vanishes if we choose
\begin{equation} 
\rho^2=
\frac{2(2\pi)^D}{g}\left[\left(\lambda-\frac{1}{\lambda}\right)^2-\mu^2\right].
\label{amplitude2} 
\end{equation} 
Then near the phase transition the amplitude of the
classical field vanishes as 
$\rho\sim \sqrt{\lambda - \lambda_c}.$  
Note that at $\mu^2=0$ $\lambda_c=1$ and $\rho\sim \lambda-\lambda_c$.

The  leading order third harmonics terms also cancel if we choose 
\begin{equation}
\Delta\phi(\mbox{\boldmath$p$}) = \rho_3\left[ e^{3i\theta_0}
\delta(\mbox{\boldmath$p$}-3\mbox{\boldmath$p$}^{(0)})+  e^{-3i\theta_0} 
\delta(\mbox{\boldmath$p$}+3\mbox{\boldmath$p$}^{(0)})\right]+O(\rho^5),
\label{correction} 
\end{equation} 
where the amplitude of the third harmonics,
$\rho_3$,
has the form 
\begin{equation} 
\rho_3=-\frac{1}{2(1-\cos
3p^{(0)})-\lambda^2 \sin^2 3p^{(0)} +\mu^2}\ \frac{g}{6(2\pi)^D}\ \rho^3 .
\label{correction2} 
\end{equation} 
It follows from (\ref{correction2}) that
$\Delta\phi(\mbox{\boldmath$p$})= O((\lambda-\lambda_0)^{3/2})$.   

In general, it is easy to see that near the phase transition
$\lambda\simeq\lambda_c$
 one can solve equation (\ref{mom_space}) iteratively,
resulting in amplitudes for the higher harmonics that have a following
behaviour near $\lambda=\lambda_c$:
\begin{equation}
\rho_{2n+1}\sim \left(\lambda-\lambda^{(0)}\right)^{(2n+1)/2}. 
\label{higher}
\end{equation}
 Here
 $\rho_{2n+1} $
 is the coefficient of 
$(2n+1)$st
 harmonics of momentum 
$\mbox{\boldmath$p$}^{(0)}$.

The conclusion is that, at least at the classical level, the system
undergoes a
second order phase transition  to a phase with spontaneously  broken
translation symmetry.  

Finally, we comment on the classical part of the energy current, 
found by substituting the classical expectation value of the field
into (\ref{energy_current}). Substituting the canonical momentum by its
lagrangian expression and using that the classical field is time
independent we obtain 
\begin{equation} \langle \int d^Dx
j_\alpha^E(\mbox{\boldmath$x$})\rangle=-\frac{\lambda^2}{2} \int d^Dx
\frac{1}{2i}\langle\left(\phi_{\mbox{\boldmath $x$}+\mbox{\boldmath
$e$}_1})-\phi_{\mbox{\boldmath $x$}-\mbox{\boldmath
$e$}_1}\right)\left(\phi_{\mbox{\boldmath $x$}+\mbox{\boldmath
$e$}_\alpha})-\phi_{\mbox{\boldmath $x$}-\mbox{\boldmath
$e$}_\alpha}\right)\rangle. 
\label{vac_density} 
\end{equation} 
This expression
is obviously zero unless 
$\alpha=1$.  
For 
$\alpha=1$ 
we obtain, after
substituting 
$\phi_{\mbox{\boldmath $x$}}=\rho e^{i\mbox{\boldmath
$p$}^{(0)}\cdot\mbox{\boldmath $x$}} $ + h.c.
\begin{equation} \langle \int d^Dx
j_\alpha^E(\mbox{\boldmath$x$})\rangle\simeq
V\rho^2\lambda^2\sin^2p_1^{(0)}\sim
V\rho^2\sim\frac{1}{g}\left\{\begin{array}{cc}\lambda-\lambda_c& {\rm if}\
\mu^2>0\\
(\lambda-\lambda_c)^2&{\rm if}\ \mu^2=0.\end{array} \right.
\label{vac_density2}  \end{equation}  
As well as this $O(1/g)$ contribution, there are perturbative quantum
corrections just as for $\lambda<\lambda_c$.

\section{Goldstone modes and the case $D=1$.}

Up to this point we have investigated the $D>1$ case only.
In one dimension, the Mermin-Wagner-Hohenberg theorem asserts that
there can be no spontaneous breaking of the continuous symmetry of
translational invariance, due to infrared singularities of the Goldstone
modes. In fact, the physics is very similar to that of spin
or charge density waves, and of modulated phases such as occur in the
ANNNI model \cite{ANNNI}, in 1+1 dimensions.

Rather than include all the fluctuations about the classical solution, it
should be adequate to derive an effective action for these Goldstone
modes only, by writing
the field $\phi(x,t)$ in the following form: 
\begin{equation} \phi(x,t) =
\rho\left\{[1+\xi(x,t)]e^{i\theta(x,t)}e^{ip^{(0)}x}\right\}+{\rm c.c.},
\label{expansion}
\end{equation}
where $\rho$ and $p^{(0)}$ are chosen to minimise the classical action,
and $\xi(x,t)$ and $\theta(x,t)$ are real. These functions are chosen so
that their fourier transforms have support only for $|p|<p^{(0)}$: in
this way the decomposition in (\ref{expansion}) is unique. In principle
we should also include higher harmonics, but, as shown in Sec.~3.3,
these are suppressed near the transition. We have included possible
longitudinal fluctuations through the field $\xi(x,t)$. 

Upon substituting  (\ref{expansion}) into the action and keeping terms
quadratic in oscillating fields $\xi(x,t)$ and $\theta(x,t)$ only we
obtain
\begin{eqnarray}
S&=&\frac{\rho^2}{(2\pi)^{D+1}}\frac{1}{2}\int d^Dp\, dE
\nonumber \\
&&\Bigl\{[D^{-1}_0(p-p^{(0)},E)+D^{-1}_0(p+p^{(0)},E)-3g\rho^2]
\,\xi(p,E)\,\xi^*(p,E) \nonumber\\&+&[D^{-1}_0(p-p^{(0)},E)+D^{-1}_0
(p+p^{(0)},E)-g\rho^2]
\,\theta(p,E)\,\theta^*(p,E)
\nonumber\\&+&i\,[D^{-1}_0(p-p^{(0)},E)-D^{-1}_0(p+p^{(0)},E)][\theta(p,E)\,\xi^*(p,E)-\theta^*(p,E)
\,\xi(p,E)]\Bigr\},
\label{action6} 
\end{eqnarray}
where the notation $D^{-1}_0(p,E)=(E +\lambda \sin p_1)^2 -2\sum_\alpha
(1-\cos
p_\alpha)-\mu^2$ has been introduced. Terms containing fields $\xi(p,E)$
or $\theta(p,E)$  taken at a  momentum of $O(p^{(0)})$ were omitted from
(\ref{action6}).  For convenience, we also symmetrised (\ref{action6}) in
momentum.

Expression (\ref{action6}) of the action has a few important features: 
\begin{itemize}
\item $p$  measures the deviation of the momentum from its ``critical''
values, $\pm p^{(0)}.$ 
\item Radial and angular degrees of freedom are mixed.  The term mixing
them is proportional
to $\lambda$.  Thus, no such term appears in the somewhat similar case of
spontaneous breaking of $U(1)$ global symmetry, where the radial and phase
degrees do  not mix at tree level. 
\item If we omit the mixing term then, as expected, the radial mode is
massive,
while the angular mode is massless,  In other words, the energy gap
vanishes at $p=0$: $D_0^{-1}(p^{(0)},0)+D_0^{-1}(-p^{(0)},0)-g\rho^2=0$ .

\end{itemize}
Thus,  due to the mixing of the two modes, which is a peculiar
feature of our model, we cannot analyse the zero modes by omitting all
degrees of freedom but the phase.
The eigenmodes of propagation are found by diagonalising the propagator
matrix in the two dimensional space $(\xi,\theta)$.
The eigenvalues of this matrix are
\begin{equation}
D^{-1}_{1,2}=\frac{\rho^2}{2}\left\{D^{-1}_0(p-p^{(0)},E)+D^{-1}_0(p+p^{(0)},E)
-2g\rho^2\pm \sqrt{
g^2\rho^4+16\lambda^2E^2\sin^2 p^{(0)}\cos^2 p}\right\}
\label{eigenvalues}
\end{equation}
In the far infrared region, 
$E,p^2<<g\rho^2$, 
one finds the mode corresponding to the
upper sign  tends to  the $\theta$-mode,
 (the $\xi$ admixture is of $O(E/g\rho^2)$), 
while the one corresponding to the lower sign tends to the
 $\xi$ mode.  Yet the dispersion relation obtained for the upper sign is
different
from the one we would have obtained, had we omitted the ``radial'' mode
from the
outset. This can be seen if we consider the asymptotic form of
eigenvalues of the inverse propagator matrix, 
\begin{equation}
D^{-1}_{1,2}=\left(\begin{array}{c}(\lambda^2-1/\lambda^2)
\left(\frac{4}{g}E^2-\rho^2p_1^2\right)-\rho^2
p_\bot^2\\
\left[\rho^2-\frac{4}{g}(\lambda^2-1/\lambda^2)\right]E^2
-(\lambda^2-1/\lambda^2)\rho^2p_1^2-\rho^2 p_\bot^2
-2g\rho^2\end{array}\right) 
\label{eigenvalues2 } 
\end{equation}
The first of these modes is massless.  The second mode appears to be
massive, except near the phase transition, if
\,$\rho^2<4(\lambda^2-1/\lambda^2)/g$, the energy squared term has the
wrong
sign. Having imaginary energy, this is not a propagating mode.  We will
soon show, however, that at small values of $\rho$ the classical approach
becomes unreliable, at least at $D=1$.

Next we shall consider the case $D=1$.  Near the phase transition the
regularised equal time correlation function of the gapless mode becomes
\begin{equation}
D(x-y) = \frac{1}{i\, 4\pi^2(\lambda^2-1/\lambda^2)}\int dE \,dp
\frac{e^{ip(x-y)}-1}{4E^2/g-\rho^2p^2}\simeq-\frac{\sqrt{g}\log|x-y|}{\,
8\pi\rho(\lambda^2-1/\lambda^2)}
\label{prop9}
\end{equation}
The correlation function is  proportional to
$1/\rho$, so it diverges as $1/\sqrt{\lambda-\lambda_c}$ near
$\lambda=\lambda_c$. 

Near $E=0$ (\ref{prop9}) approximates the correlation function for the
phase mode of
oscillations.  Then the equal time correlation function of
$\phi(x,0)$ is
\begin{equation}\langle\phi(x,0)\phi(y,0)\rangle\sim
\rho^2\langle e^{i(\theta(x,0)-i\theta(y,0)+ip^{(0)}(x-y)}\rangle \sim
\frac{\rho^2\cos[ p^{(0)}(x-y)]}{|x-y|^\eta},
\label{correlation}
\end{equation}
where $\eta=\sqrt{g}/[8\pi \rho (\lambda^2-1/\lambda^2)]$,
which exhibits
an oscillating power-like behaviour with a nonuniversal critical exponent.
This means that there is only quasi-long range order in $D=1$. 

Note that the parameter $\rho\propto(\lambda-\lambda_0)^{1/2}$ 
plays a very similar role to $\beta=1/T$ in
the $XY$-model.  Just as in this model, there are also vortex-like
excitations whose condensation at high enough $T$ (small $\rho$)
completely destroys even the quasi-long range order. In our case,
these result from the fact that $\theta(x,t)$ is an angular variable,
defined only modulo $2\pi$. A `vortex' corresponds to a fluctuation in
which an extra oscillation of the field $\phi(x,t)$, localised in $x$,
is either inserted or removed in a large but finite space and
time interval. This has
an action which grows like the logarithm of the system size. By the
Kosterlitz-Thouless criterion~\cite{KT}, they become relevant when
$\eta>1/4$. 

We conclude that the transition to the quasi-modulated phase occurs not
at $\lambda\approx\lambda_0$, but at a higher value when $\rho$ has
become sufficiently small. At the transition, we should then have
$\eta=1/4$, with the value of this exponent decreasing as we go further
into the modulated phase.  

Although this phase exhibits only quasi-long range order, it nevertheless
supports a non-zero classical current, as for $D>1$, since the leading
contribution $\sim \lambda\rho^2\sin^2 p^{(0)}$ is not affected by
infrared problems. In fact, this quantity corresponds to the superfluid
density in the XY model, and is expected to jump discontinuously at the
transition~\cite{Nelson}.

In the ultraviolet limit of (\ref{eigenvalues}) the two roots of the
inverse
propagator reduce to the two poles of (\ref{prop3}).  This shows the
consistency of our descriptions of the critical and broken phases.

\section{Energy current in the ferromagnetically ordered state.}

In the regime $\mu^2\leq0$ we have not been able to obtain
results comparable to those above, especially in the
difficult case of $\mu^2=0$. We can show however, at least at the
classical level, that the second order transition at  $\mu^2>0$ turns into
a first order transition at $\mu^2<0$. 

When $\mu^2<0$ the $Z_2$
symmetry of the theory is broken even  at $\lambda\ll1$, where the
translation invariance is certainly not broken. 
Thus, we must shift the field to
minimise the effective potential, but the classical
field is  a constant, $\phi_0$.  When $\lambda$ is increased
the classical field will  become coordinate dependent above a certain
critical value.  Writing the classical
field as $\phi(p_1)=\phi_0+\psi(p_1)$ the extremum condition for field
$\psi$ will read as: 
\begin{eqnarray} i\frac{\delta \Gamma_{\rm
eff}}{\delta\psi(\mbox{\boldmath$p$})
}=0&=&\left[\lambda^2\sin^2p_1-2\sum_\alpha^D(1-\cos p_\alpha)
-2|\mu^2|\right]\psi( \mbox{\boldmath$p$}) \nonumber \\
&-&\frac{g}{2}\frac{\phi_0}{(2\pi)^{D}}\int d^Dp_1d^Dp_2 \, \psi(
\mbox{\boldmath$p$}_1)\psi(
\mbox{\boldmath$p$}_2)\delta(\mbox{\boldmath$p$}-
\mbox{\boldmath$p$}_1-\mbox{\boldmath$p$}_2
)\nonumber \\ &-&\frac{g}{6}\frac{1}{(2\pi)^{D}}\int d^Dp_1d^Dp_2d^Dp_3 \, 
\psi(
\mbox{\boldmath$p$}_1)\psi( \mbox{\boldmath$p$}_2)\psi(
\mbox{\boldmath$p$}_3)\delta(\mbox{\boldmath$p$}-\mbox{\boldmath$p$}_1-
\mbox{\boldmath$p$}_2
-\mbox{\boldmath$p$}_3), 
\label{mom_space4} 
\end{eqnarray}
 where
$\phi_0=\sqrt{6|\mu^2|(2\pi)^D/g}$.

We can rescale (\ref{mom_space4}) by dividing by $|\mu^2|\phi_0$ and 
introducing the new field $\psi\rightarrow \psi \phi_0$.  Then, using the
fact that the ground state depends only on $p_1$ (in what follows we will
drop the subscript) , one obtains the following equation for $\psi$:
\begin{eqnarray}
0&=&\left[2-\frac{\lambda^2\sin^2p-2(1-\cos p)}{|\mu^2|}
\right]\psi( p) \nonumber \\
&+&3\int dp^1dp^2 \, \psi(
p^1)\psi(
p^2)\delta(p-
p^1-p^2
)\nonumber \\ &+&\int dp^1dp^2dp^3 \, 
\psi(
p^1)\psi( p^2)\psi(
p^3)\delta(p-p^1-p^2-p^3). 
\label{mom_space5} 
\end{eqnarray}

Naively, one would think that just like for
$\mu^2>0$, 
at least at the
tree level, the phase transition is determined by the linear term
in (\ref{mom_space5}) and occurs at
$\lambda=\lambda_1$, 
where
$\lambda_0-1/\lambda_0=\sqrt{2}\mu$. (Note the extra factor of
$\sqrt{2}$, reflecting the fact that the bare mass for $\mu^2<0$ 
is $\sqrt2|\mu|$.) 
Contrary to this expectation we will
prove that, due to the presence of the term quadratic in $\psi$ in
(\ref{mom_space5}),
the system undergoes a  {\em first order} phase
transition, at
a critical value of $\lambda=\lambda_c<\lambda_1$, 
to a state with broken  translation invariance.  
In order to show this we merely have to exhibit a solution of 
(\ref{mom_space5}) which gives a lower action, for
$\lambda_c<\lambda<\lambda_1$, than the solution $\psi(p)=0$. To this
end, consider the ansatz
\begin{eqnarray} \psi(p_1)&=&\psi_0 \delta(p_1) + \rho_1
\left[e^{i\theta_0}\delta(p_1-p_1^{(0)})+e^{-i\theta_0}
\delta(p_1+p_1^{(0)})\right]\nonumber \\ &+&
\psi_2\left[e^{i2\theta_0}\delta(p_1-2p_1^{(0)})
+e^{-i2\theta_0}\delta(p_1+2p_1^{(0)})\right]
 \label{ansatz4} 
\end{eqnarray}
where, as before, $\cos p^{(0)}=1/\lambda^2.$
After substituting  this
ansatz into (\ref{mom_space4}) we minimise the action with respect to 
$\psi_0$, $\rho_1$, and $\psi_2$.  We obtain  
\begin{eqnarray}
-2\psi_0&=&3(\psi_0^2+2\rho_1^2
+2\psi_2^2)+\psi_0^3+3\psi_0(\rho_1^2+\psi_1^2)\nonumber \\
\epsilon\rho_1&=&
6(\psi_0\rho_1+\psi_2\rho_1)+3\rho_1(\rho_1^2
+\psi_0^2+2\psi_2^2)\nonumber\\
-2\psi_2 &=&
6(\rho_1^2+\psi_2\psi_0)+
2\psi_0\rho_1^2+\psi_0^2\psi_2
\label{conditions} 
\end{eqnarray} where we used the relation
$\lambda^2\sin^2(2p_1^{(0)})-2[1-\cos(2 p_1^{(0)})]=0$ and the notation
${\left(\lambda-\frac{1}{\lambda}\right)^2}/{|\mu^2|}-2=\epsilon$.

Note that in terms of  the new variables a second order transition is
possible
only at $\epsilon\geq0$. (\ref{conditions})  can however be solved
numerically
and doing so we find that it has a real solution for all\,
$\epsilon>\epsilon_0=-0.3395$.  This value of $\epsilon$ corresponds to
$\lambda_c\simeq 1.834 <\lambda_1=1.932$.  Below $\lambda_c=1.834$ there
is no
real solution of the form (\ref{ansatz4}), but
above this
value a pair of complex roots become real and the parameters take the
values
$\rho_1\simeq 0.130$, $\psi_0\simeq- 0.0410$, and $\psi_2=-0.0358$. The
true
tree level transition is then at a value $\lambda_c< 1.834.$   The phase
transition line is at a fixed value of $(\lambda-1/\lambda)^2/|\mu^2|.$
This
fixed value is smaller than 1.6605, and certainly smaller than the
superficial
prediction, 2, that was based on the assumption that the transition is
second order.  
 
Of course, this argument does not imply that the ansatz (\ref{ansatz4})
represents the true minimum of the effective potential (now
there is no reason to ignore higher harmonics as there was in the
second order case) but is does imply that the transition cannot be
second order, at least at tree level. Nor does the argument rule out the
possibility of a restoration of the second order nature of the
transition by the fluctuations, especially in low dimensions (see next
section). 
\section{Summary}

Following investigations of similar nature for integrable one dimensional
spin chains~\cite{antal1}~\cite{antal2}, we have studied the effect of
an energy current in field theories, at zero temperature. 
The energy current was imposed on the system by
adding the global energy current operator, multiplied by a Lagrange
multiplier, $\lambda$, to the Hamiltonian.  The common feature of these
models is the appearance of a phase transition at a critical value of
$\lambda$. 

Field theoretic models modified by an energy current can roughly be
divided into two groups.   In the first, less interesting group of
theories the hamiltonian is destabilised at the critical value of
$\lambda$, $\lambda_c$, whereby the energy spectrum has no lower bound at
$\lambda>\lambda_c$. A typical example for these models is a relativistic,
self-interacting, real, scalar field theory.  Theories, in which at $D>1$
(where $D$ is the number of spatial dimensions) translation invariance
breaks spontaneously, are more interesting.  A typical example for this
second group is a self-interacting real scalar field theory on a lattice. 

 In the lattice models that we consider the global energy
current does not commute with the Hamiltonian and consequently, unlike in
spin models~\cite{antal1}~\cite{antal2}, the ground state expectation
value of the energy current  has a nonvanishing perturbative contribution
at all  $\lambda\not=0$. Another, nonperturbative contribution to the
energy current appears at the phase transition point.

 The system has a few unusual features in the broken phase, at $D>1$. The
first of these is that the field  has a nonvanishing, oscillating vacuum
expectation value. Correlation functions are also oscillating functions of
the
coordinate. 
The oscillations are the consequence of the unusual form of the
energy-momentum dispersion relation and the vanishing of the energy gap at
a nonvanishing value of the momentum.   The second order phase transition
to the broken phase is accompanied by the appearance of a Goldstone boson,
in spite of the fact that the translation symmetry is a discrete symmetry
on the lattice. This happens because near the phase transition the lattice
structure of the system becomes irrelevant. 

The second interesting feature of the field theoretic model is the
apparent spontaneous generation of degrees of freedom and enlargement of
the internal symmetry near  the phase transition. As the energy gap
vanishes at $p=p^{0}\not=0$, it is meaningful to rewrite the real scalar
field into the form $\phi(x)=\chi(x)e^{ip^{(0)}x}+\chi^\dagger(x)
e^{-ip^{(0)}}$. At least in the infrared domain ($p<<p^{(0)}$) $\chi(x)$
is a genuine complex field with $U(1)$ symmetry (planar model) in contrast
to the $Z_2$ symmetry of the of the field $\phi(x)$.  When the vacuum
expectation value of  $\chi$ is non-vanishing, its phase degree of freedom
becomes a Goldstone mode.  The extra near infrared degrees of freedom can
of  course be related to large momentum modes of the original theory, so
in reality the number  of degrees of freedom is unchanged. What we really
observe is the transmutation of a space-time symmetry (translation
invariance) into a broken internal symmetry (phase transformations of
field $\chi$). 

The analogy with the $U(1)$ model in $D>1$ spatial dimensions is not
complete. Due to the peculiar form of the single particle dispersion
relation the propagator takes a nonrelativistic form in the infrared
domain. This leads to a critical dynamics different than that of a complex
scalar field, and the anomalous dimension vanishes at all $D>1$.

When $D=1$ the translation invariance cannot be broken
spontaneously. Yet the characteristic momentum $p^{(0)}$ and the
complexification of the degrees of freedom survive
due to large fluctuations,  the field $\chi(x)$ must have a vanishing
position-dependent
vacuum expectation value.  Still, the analogy with the $U(1)$ symmetric
$XY$ model, based on the decomposition of the scalar field into a
superposition of a complex field and of its complex conjugate, is
maintained.  One obtains damped oscillating correlation functions with
non-universal anomalous dimensions. The phase transition at
$\lambda=\lambda_c$ is driven by the condensation of vortices of the
$\chi(x)$ field. 

Despite some similarities with the exactly solvable cases treated in
\cite{antal1}~\cite{antal2}, there are also some important differences
in our results. Partly this is due to the fact that in our case the
current does not commute with the hamiltonian, so that the current is
non-vanishing even for arbitrarily small bias $\lambda$. However, the
singular part of the current, which we have argued occurs only in its
`classical' piece, has a different behaviour, vanishing linearly as
$\lambda\to\lambda_c+$, rather than proportional to
$\sqrt{\lambda-\lambda_c}$ as in \cite{antal1}. In $D=1$, this linear
vanishing is modified by the vortices to a discontinuous jump. This
behaviour is related to that of the correlation function: we find that
in our model it should exhibit oscillations modulated by a power law
$|x|^{-\eta}$, where $\eta$ is continuously varying throughout the
modulated phase, increasing to the universal value $1/4$ at the
transition. This should be compared with the constant value $\eta=1/2$
found for the transverse Ising chain in \cite{antal1}. 

These differences in detail are not necessarily surprising. When such
integrable models as the Ising or XY chains are mapped onto coulomb gas
models, the additional conservation laws in the integrable system may
show up in unusual constraints (for example the charge at infinity)
which can modify the critical exponents away from their usual values.
However it should be clear that our results do not depend so much on
details of our model, and therefore should possibly be more generic.

A comment should be made concerning the time-dependence in our model. In
most of this paper the parameter $t$ referred to as `time' is merely an
artifice introduced in order to provide a lagrangian formulation of
the problem of finding the ground state properties of the operator
$H^{(0)}-\lambda J$. Thus no meaning should be attached to quantities
which are not $t$-independent. The true time dependence is still
generated by $H^{(0)}$. In particular, this means that the expectation
value of the current, which does not commute with 
$H^{(0)}$, should be time-dependent. The constant quantity which we
have computed is therefore only its zero-frequency part. It would be
interesting, but difficult, to investigate the time-dependent part as
well.

In fact, the mathematics of our model bears a very close resemblance to
that which appears in the study of other systems which exhibit spatially
modulated phases, such as spin or charge density waves, or
systems such as the ANNNI model \cite{ANNNI}
with competing ferromagnetic nearest and
antiferromagnetic next-nearest neighbour interactions. In fact, if one
neglects the `time'-dependence in (\ref{lagrange}), and takes the limit
$g\to\infty$, $\mu^2\to-\infty$ with $\mu^2/g$ fixed, so that the field at
each site is
frozen to the values $\pm\sqrt{3|\mu^2|/g}$, this becomes precisely the
energy function for the classical ANNNI model in $D+1$ dimensions. Even
in mean field theory
the phase diagram for this model is extremely
rich. Between the ferromagnetic $\cdots++++\cdots$ phase and 
the $\cdots++--++\cdots$ antiphase lie an infinite number of
other modulated phases. In 1+1 dimensions, these can also be separated
by Kosterlitz-Thouless incommensurate phases, bounded by
Pokrovsky-Talapov transitions\cite{PT}. It should be stressed that the
limit 
above is quite different from the regime considered in the body of this
paper where it was assumed that the characteristic wave vector $p^{(0)}$
is much smaller than the reciprocal lattice vector, and that lattice
effects are irrelevant. The modulated phases considered there are
therefore incommensurate. The `time'-dependent fluctuations in our model
are rather different from those in transverse direction of the
ANNNI-type models. Nevertheless it is possible that for $D=1$ the
first-order transition predicted in the last section is replaced by one
of Pokrovsky-Talapov type. 

Besides these correspondences, there remains the important question of
the relevance of these kinds of model, whether integrable as in
Refs.~\cite{antal1},\cite{antal2}, or non-integrable as in our example,
to the study of non-equilibrium processes at finite temperature where
dissipation plays a crucial role. It is to be hoped that some of the
phenomena discussed in this paper will survive into this regime.
\vskip0.1in

\noindent{\bf Acknowledgements}\\
The authors wish to thank T.~Antal, F.~Essler, Z.~R\'acz and J.~Yeomans for
discussions. 
 J.C. was supported in part by the Engineering and
Physical
Sciences Research Council under Grant GR/J78327 while P.S. by the U.S. Department of Energy under grant
number DE-FG02-84ER-40153.


\begin{thebibliography}{10} \small \addtolength{\itemsep}{-6pt} 
\bibitem{first} M.C. Cross and P.C. Hohenberg, Rev. Mod. Phys. {\bf 65}
(1994) 851.
\bibitem{second} B. Schmittmann and R. P. K. Zia, in {\em Phase
Transitions and Critical Phenomena}, Eds. C. Domb and J.L. Lebowitz
(Academic Press, New York, 1995).
\bibitem{third} S. Katz, J.L. Lebowitz, and H. Spohn, J. Stat. Phys. {\bf
34} (1984) 497; A. DeMasi, P.A. Ferrari. and J.L. Lebowitz, Phys. Rev.
Lett. {\bf 55} (1985) 1947.  
\bibitem{fourth} T.L. Hill, J. Chem. Phys. {\bf 76} (1982) 1122; P.L.
Garrido, A. Labarta, and J. Marro, J. Stat. Phys. {\bf 49} (1987) 551.
 \bibitem{antal1} Antal, Z. R\'acz, and  Sasv\'ari,
Phys. Rev. Lett. {\bf 78} (1997)  167. 
\bibitem{antal2} T. Antal, Z. R\'acz, A.
R\'akos, and G. M. Sch\"utz, Phys.Rev. E {\bf 57} (1998) 5184.
\bibitem{jona} G. Jona-Lasinio, Nuovo Cim. {\bf 34A} (1964) 1790.
\bibitem{peskin}  M. Peskin and  Schroeder, ``An Introduction to Quantum
Field
Theory,''  Addison-Wesley (Reading, Mass. 1995).
\bibitem{ANNNI} See, for example, J.~Yeomans, Solid State Physics {\bf
41} (1988) 151; W.~Selke, Phys. Rep. {\bf 170} (1988) 213.
\bibitem{KT} J.~M.~Kosterlitz and D.~J.~Thouless, J. Phys. C {\bf 5}
(1972) L124; J. Phys. C {\bf 6} (1973) 1181.
\bibitem{Nelson} D.~R.~Nelson and J.~M.~Kosterlitz, Phys. Rev. Lett.
{\bf 39} (1977) 1201.
\bibitem{PT} V.~L.~Pokrovsky and A.~L.~Talapov, Phys. Rev. Lett.
{\bf 42} (1979) 66.

 \end{thebibliography}
 \end{document}